\documentstyle[12pt]{article}

\newcommand{\dd}{{\rm d}}                                
\newcommand{\ddd}[1]{{\frac{\rm d}{{\rm d}#1}}}          
\newcommand{\vdd}[1]{{\frac{\delta }{\delta #1}}}        
\newcommand{\be}{\begin{equation}}                                  %
\newcommand{\ee}{\end{equation}}                                    %
\newcommand{\ba}{\begin{eqnarray}}                                  %
\newcommand{\ea}{\end{eqnarray}}                                    %
\newlength{\ypit}                                 
\newcommand{\s}[1]{                                                 %
\setlength{\ypit}{\unitlength}  				    %
\settowidth{\unitlength}{$#1$}  				    %
\addtolength{\unitlength}{-2mm}                                     %
\mbox{ 								    %
\begin{picture}(1,1) 				                    %
\put(0,0){\makebox(1,1)[b]{$#1$}} 				    %
\put(0,0){\makebox(1,1)[b]{/}}					    %
\end{picture} 							    %
}								    %
\setlength{\unitlength}{\ypit} 					    %
}								    %

\begin{document}


\begin{titlepage}
\begin{flushright}
UU-ITP 30/93\\
hep-th/9311110
\end{flushright}
\large
\vspace*{25mm}
\begin{center}
{\LARGE\rm\bf Symplectic Geometry of Supersymmetry and Nonlinear
Sigma
Model}
\vspace*{10mm}

{\large K. Palo\\} {\small\it\noindent Department of Theoretical
Physics, Uppsala University \\ Box 803, S-75108 Uppsala, Sweden\\
and\\ Institute of Physics, Estonian Academy of Sciences\\ 142 Riia
Str., EE-2400 Tartu, Estonia\\}
\vspace*{20mm}

\end{center}
\begin{normalsize}
Recently it has been argued, that Poincar\'{e} supersymmetric field
theories admit an underlying loop space hamiltonian (symplectic)
structure.  Here shall establish this at the level of a general $N=1$
supermultiplet. In particular, we advocate the use of a superloop
space introduced in \cite{mnp2}, and the necessity of using
nonconventional auxiliary fields. As an example we consider the
nonlinear $\sigma$-model. Due to the quartic fermionic term, we
conclude that the use of superloop space variables is necessary for
the action to have a hamiltonian loop space interpretation.
\end{normalsize}


\vfill
\noindent\rule{\textwidth}{0.5mm}
\\
\normalsize
\noindent E-mail: palo@rhea.teorfys.uu.se
\normalsize\rm
\end{titlepage}


{\bf 1.  Introduction.}  Recently, a conceptually new approach has
been developed to describe Poincar\'e supersymmetric field
theories. In this approach supersymmetric theories are interpreted in
terms of loop space symplectic geometry \cite{mnp1,mnp2}.  The idea
originates from Witten, and was presented by Atiyah in
\cite{atiyah}. He considered the path integral for
a supersymmetric spinning particle in a gravitational background, to
derive the Atiyah-Singer index theorem for a Dirac operator on a
Riemannian manifold. He argued that the path integral admits an
underlying loop space symplectic structure such that the relevant
Hamiltonian flow is integrable. As a consequence, he could then
evaluate the path integral by localization methods, using an
infinite dimensional analog of the Duistermaat-Heckman integration
formula
\cite{duihec}. A more detailed mathematical investigation
was subsequently discussed by Bismut \cite{bismut}, who also
discussed some generalizations.

An approach to include an arbitrary gauge field background was
considered in \cite{hmnp}. Subsequently in
\cite{mnp1, mnp2} it was argued by explicit analysis of
various examples, that a loop space symplectic structure is not just
a property of the particular model but is rather a characteristic
feature of generic supersymmetric theories.

Here we shall consider the symplectic interpretation of generic
$N=1$ Poincare supersymmetric theories, at the level of a general
supermultiplet. We explain the hamiltonian loop space construction
in a model independent fashion, and argue in favor of a superloop
space and a new auxiliary field construction on geometrical
grounds. In particular, we show how our construction applies to the
supersymmetric nonlinear sigma model, with a quartic fermionic
self-interaction. A conventional, bosonic loop space construction
does not apply in this case, since a four-form does not admit any
natural geometric interpretation in terms of symplectic
geometry. However, in the {\it superloop} space the quartic fermion
term turns out to be just a symplectic two-form.


\vspace*{10mm}
{ \bf 2. Superloop space symplectic geometry.} We shall consider a
supersymmetric field theory, with a generic (bosonic or fermionic)
field $\Phi(x,t)$ vanishing in the spatial infinity and periodic in
time:
\begin{eqnarray}
\Phi(x,t)  \rightarrow  0, \;{\rm if}\; x\rightarrow\infty\, ,
\nonumber\\
\Phi(x,t) = \Phi(x,t+T)\label{boundary}\,.
\end{eqnarray}
With these boundary conditions, we can view the fields as defined on
a loop space. As we shall see, in a supersymmetric theory the fields
can be naturally divided into two different categories: Half of the
fields are interpreted as loop space coordinates, and the other half
as the corresponding loop space one-forms.  We denote these fields
as $\phi(x,t)$ and $\xi(x,t)\,\sim\,\delta \phi(x,t)$ respectively,
and we emphasize that we do not necessarily identify bosonic fields
as coordinates and fermionic as one-forms: The loop space can have
both bosonic and fermionic coordinates, and the corresponding
one-forms are then fermionic and bosonic respectively. In
particular, functionals of the original fields are now loop space
differential forms.  We define the loop space exterior derivative
\begin{equation}
{\rm d}=\int{\rm d} x\oint{\rm d} t\, \xi(x,t)\vdd{\phi(x,t)}\,.
\label{deriv}
\end{equation}
and inner multiplication by introducing a preferred vector field: On
the loop space, there is a natural family of vector fields --- the
time-like derivatives of coordinates. However, in the following we
shall find it more convenient to consider loops parametrized by
light-cone coordinates, and thus we consider inner multiplication by
vector fields
\begin{equation}
i_{X}=\int{\rm d} x\oint{\rm d} t\,
\partial_{\tau}\phi(x,t)\vdd{\xi(x,t)}\,,
\label{ichi}
\end{equation}
where $\tau$ denotes one of the light-cone variables $x\pm t$.  We
shall prove that for a generic $N=1$ supersymmetric field theory the
generators of Poincar\'{e} supersymmetry transformations can be
expressed as
\begin{equation}
Q={\rm d}+i_{X}
\label{Q}
\end{equation}
with respect to the light-cone vector fields in the various
light-cone directions. In refs. \cite{mnp1,mnp2}, this was already
shown to be the case in various examples.

The square of $Q$ admits a definite geometric meaning:
\begin{eqnarray}
Q^{2} & = & ( {\rm d}+i_{X})^{2}= \nonumber \\
      & = & {\rm d}i_{X}+i_{X}{\rm d}={\cal L}_{X}\, ,
\label{Lie}
\end{eqnarray}
is the Weyl formula for Lie derivative in the light-cone direction.
Making use of eqs.\,(\ref{deriv}) and (\ref{ichi}) one then obtains
\begin{equation}
Q^{2}\sim\frac{\partial \phi}{\partial \tau}\vdd{\phi}+
\frac{\partial \xi}{\partial \tau}\vdd{\xi}
\equiv\frac{\partial}{\partial \tau }\,.
\label{Q2}
\end{equation}
(Here and afterwards integration over space-time is assumed without
writing it explicitly.) As we will see below eq. (\ref{Q2}) is a
representation of supersymmetry algebra.

We shall find, that the action of a supersymmetric model can be
naturally divided divided into a sum of a loop space scalar $\cal H$
and a loop space two form $\omega$:
\begin{equation}
{\cal S}={\cal H}+\omega\, ,
\label{action}
\end{equation}
For explicit examples see below and refs.\,\cite{mnp1,mnp2}.

Due to our boundary conditions the space-time integrals of total
derivatives vanish:
\begin{equation}
\int{\rm d} x\oint{\rm d} t\,\partial_{\tau}F(\phi,\xi)=0\,.
\label{TotDer}
\end{equation}
Hence (\ref{Q2}) actually becomes
\begin{equation}
Q^{2}=0\,.
\end{equation}

When we investigate the consequences of supersymmetry of the action
--- $Q{\cal S}=0$ --- we discover the following equations by
separating differential forms of different degrees:
\begin{eqnarray}
{\rm d}\omega= 0\, , \label{symp} \\ {\rm d}{\cal H}+i_{X}\omega=0\,
.\label{hameq}
\end{eqnarray}
(\ref{symp}) implies that $\omega$ can be interpreted as a
symplectic two-form and from (\ref{hameq}) we conclude that $X$ is
the Hamiltonian vector field corresponding to ${\cal H}$. The
concrete form of $X$ is determined by the supersymmetry algebra in a
model-independent way. As $X$ has a very simple structure
(\ref{ichi}), we can integrate the corresponding "Hamiltonian
equations of motion" in the loop space getting constant modes as the
solution. From the point of view of Duistermaat-Heckman integration
formula, this means that the path integral corresponding to a
supersymmetric action is localized to constant modes.  (See
refs. \cite{mnp1, mnp2, hmnp} for discussion.)

We also observe, that ${\cal H}$ is uniquely (up to a total
derivative and a constant) determined by $\omega$.  For a given
$\omega$ one can locally find the corresponding symplectic potential
$\vartheta$ that fulfills the following condition:
\begin{equation}
{\rm d}\vartheta=\omega\, .
\label{potent}
\end{equation}
Acting with $Q^{2}$ on $\vartheta$ one gets
\begin{equation}
{\rm d}(i_{X}\vartheta)+i_{X}\omega=0\, ,
\label{MomMap}
\end{equation}
and taking into account eq. (\ref{hameq}) we identify:
\be
i_{X}\vartheta={\cal H}\,.
\label{BlauCondition}
\ee
In order to
establish uniqueness of the choice of ${\cal H}$ let us choose
another potential
$\vartheta'=\vartheta+{\rm d}\varphi$ for some scalar $\varphi$.
We get
\begin{eqnarray}
i_{X}\vartheta' & = & i_{X}\vartheta + i_{X}{\rm d}\varphi=
\nonumber \\
& = & i_{X}\vartheta + \ddd{\tau}\varphi={\cal H}\, .
\end{eqnarray}
Here we used (\ref{TotDer}) to put $\dot \phi = 0$.  On the other
hand one might assume, that the true Hamiltonian ${\cal H}'$ differs
from $i_{X}\vartheta$. However, it follows from the supersymmetry
(\ref{hameq}), that for ${\cal H}=i_{X}\vartheta$ and ${\cal H}'$
one has
\begin{equation}
{\rm d}({\cal H}-{\cal H}')=0\, ,
\end{equation}
Thus, modulo a total derivative ${\cal H}$ and ${\cal H}'$ can differ
only by a constant mode, and from (\ref{MomMap}) and
(\ref{BlauCondition}) we can locally write the action as a
supersymmetry variation,
\begin{equation}
{\cal S}=({\rm d}+i_{X})\vartheta\, .
\end{equation}


\vspace*{10mm}

{\bf 3. Explicit constructions.}  In this section we shall
explicitly realize the geometrical structures of the previous
section using $N=1$ super-Poincar\'{e} algebra in four dimensions
(see e.g. \cite{sohnius}):
\begin{equation}
\{Q_{\alpha},Q_{\beta}\}=2(\gamma^{\mu}{\cal C})P_{\mu}\, .
\label{salg1}
\end{equation}
We use a Majorana representation with
$\gamma^{0}=-\sigma^{2}\otimes I,\,\,
\gamma^{1}=-i\sigma^{3}\otimes\sigma^{1},\,\,
\gamma^{2}=i\sigma^{1}\otimes I,\,\,
\gamma^{3}=-i\sigma^{3}\otimes\sigma^{3},$
where we have:
\begin{equation}
(\gamma^{\mu}{\cal C})P_{\mu}= \left(\begin{array}[c]{cccc}
i\partial_{+} & \ast & \ast & \ast \\ \ast &i
\partial_{+} & \ast & \ast
\\ \ast & \ast &i\partial_{-} & \ast \\ \ast & \ast & \ast
&i\partial_{-}
\end{array}     \right)\, ,
\label{salg2}
\end{equation}
with light-cone derivatives on the diagonal and $\ast$ standing for
terms that are not relevant in the following.  Eq. (\ref{salg1})
suggests that $Q$ --- given by eqs.\,(\ref{deriv}\,--\,\ref{Q}) and
satisfying (\ref{Q2}) --- can be identified with any of the
$Q_{\alpha}$-s, where light-cone coordinates $x^{\pm}=x_2 \pm t$
stand for the parameter $\tau$. Different representations for gamma
matrices would define different preferred lightcone directions.

In order to demonstrate that the symplectic structure is present for
a general case it is sufficient to prove it for the general $N=1$
supermultiplet \cite{sohnius} containing a complex scalar $M$,
pseudoscalars $C,\,N,\,D$, and a vector $A_{\mu}$, and two Dirac
spinors $\chi$ and $\lambda$. Other multiplets can be obtained by
imposing some additional constraints.

{}From the transformation rules \cite{sohnius} of the complex $N=1$
supermultiplet $V=(C;\,\chi;\,M,\,N,\,A_{\mu};\,\lambda;D)$:
\begin{eqnarray}
\delta C &= & \bar{\zeta}\gamma_{5}\chi\, ,        \nonumber\\
\delta\chi & = & (M+\gamma_{5}N)\zeta-i\gamma^{\mu}
(A_{\mu}+\gamma^{5}\partial_{\mu} C)\zeta\, , \nonumber\\
\delta M & = & \bar{\zeta}(\lambda-i\s\partial\chi)\, ,
\nonumber\\
\delta N & = & \bar{\zeta}\gamma_{5} (\lambda-i\s
\partial \chi)\, ,\nonumber\\
\delta A_{\mu} & = & \bar{\zeta}
(i\gamma_{\mu}\lambda+\partial_{\mu}\chi)\, , \label{trans0} \\
\delta \lambda & = & -i\sigma^{\mu\nu}\zeta\partial_{\mu}A_{\nu}-
\gamma_{5}\zeta D\, , \nonumber\\
\delta D & = & -i\bar{\zeta}\s\partial\gamma_{5}\lambda\, ,
\nonumber
\end{eqnarray}
we can find the transformation generated by any of the
$Q_{\alpha}$-s.  For $Q_{1}$ we get:
\begin{eqnarray}
Q_{1} C & = & \chi_{2}\, ,
\nonumber\\ Q_{1} \chi_{1} & = & iA_{+}\, ,
\nonumber\\ Q_{1} \chi_{2} & = & i\partial_{+}C\, ,
\nonumber\\ Q_{1}
\chi_{3} & = & i(M+A_{z}+\partial_{x}C)\, , \nonumber\\ Q_{1}
\chi_{4} & = & i(N+A_{x}-\partial_{z}C)\, , \nonumber\\ Q_{1}
M & = & \lambda_{1}+\partial_{+}\chi_{3}+\partial_{x}\chi_{2}-
\partial_{z}\chi_{1}\, , \nonumber\\ Q_{1}
N & = & \lambda_{2}+\partial_{+}\chi_{4}-\partial_{x}\chi_{1}+
\partial_{z}\chi_{2}\, , \nonumber\\ Q_{1}
A_{+} & = & \partial_{+}\chi_{1}\, , \label{trans1} \\ Q_{1}
A_{-} & = & -2\lambda_{3}+\partial_{-}\chi_{1}\, , \nonumber\\ Q_{1}
A_{x} & = & -\lambda_{2}+\partial_{1}\chi_{1}\, , \nonumber\\ Q_{1}
A_{z} & = & -\lambda_{1}+\partial_{z}\chi_{1}\, , \nonumber\\
Q_{1}\lambda_{1} & = & -i(\partial_{+}A_{z}-\partial_{z}A_{+})\, ,
\nonumber\\ Q_{1}\lambda_{2} & = & -i(\partial_{+}A_{x}-
\partial_{x}A_{+})
\,, \nonumber\\ Q_{1}\lambda_{3} & = & -\frac{i}{2}
(\partial_{+}A_{-}-
\partial_{-}A_{+})\, , \nonumber\\
Q_{1}\lambda_{4} & = & -i(D+\partial_{x}A_{z}-\partial_{z}A_{x})\,
,\nonumber\\ Q_{1} D & = & -\partial_{+}
\lambda_{4}-\partial_{z}\lambda_{2}+
\partial_{x}\lambda_{1}\, .  \nonumber
\end{eqnarray}
To obtain notational simplicity we use redefined fields denoted by
primes:
\begin{eqnarray}
M' & = & M+A_{z}+\partial_{x}C\, , \nonumber\\
N' & = & N+A_{x}-\partial_{z}C\, , \nonumber\\
\lambda_{1}' & = & \lambda_{1}-\partial_{z}\chi_{1}\, , \nonumber\\
\lambda_{2}' & = & \lambda_{2}-
\partial_{x}\chi_{1}\, , \label{redef}\\
\lambda_{3}' & = & 2\lambda_{3}-\partial_{-}\chi_{1}\, , \nonumber\\
D' & = & D+\partial_{x}A_{z}-\partial_{z}A_{x}\, , \nonumber
\end{eqnarray}
and thus we can rewrite (\ref{trans1}) in a more compact form:
\begin{eqnarray}
Q_{1} C & = & \chi_{2}\, ,\nonumber\\
Q_{1} (\chi_{1}, \chi_{2}, \chi_{3},
\chi_{4}) & = &  (iA_{+}, i\partial_{+}C, iM', iN')\, ,\nonumber\\
Q_{1} M' & = & \partial_{+}\chi_{3}\, , \nonumber\\
Q_{1} N' & = & \partial_{+}\chi_{4}\, , \label{trans2}\\
Q_{1} (A_{+}, A_{-}, A_{x},
A_{z}) & = &  (\partial_{+}\chi_{1}, -\lambda_{3}, -\lambda_{2},
-\lambda_{1})\, ,
\nonumber\\
Q_{1} (\lambda_{1}', \lambda_{2}', \lambda_{3}', \lambda_{4}) & = &
(-i\partial_{+}A_{z}, -i\partial_{+}A_{x}, -i\partial_{+}A_{-}, -iD')
\, ,\nonumber\\
Q_{1} D' & = & -\partial_{+}\lambda_{4}\, .\nonumber
\end{eqnarray}
Equations (\ref{redef}) are exactly the definitions of the
nonstandard auxiliary fields, that were introduced in
\cite{mnp1,mnp2}.  Eqs. (\ref{trans2}) suggest us to write:
\begin{eqnarray}
{\rm d} & = & \chi_{2}\vdd{C}+iA_{+}
\vdd{\chi_{1}}+iM'\vdd{\chi_{3}}+
iN'\vdd{\chi_{4}}+ \nonumber\\  &   & -\lambda_{3}'\vdd{A_{-}}-
\lambda_{2}'\vdd{A_{x}}-\lambda_{1}'\vdd{A_{z}}-
iD'\vdd{\lambda_{4}}\, ,
\label{reald}
\end{eqnarray}
and
\begin{eqnarray}
i_{X_{+}}  & = &  i\partial_{+}C\vdd{\chi_{2}}+
\partial_{+}\chi_{3}\vdd{M'}
+\partial_{+}\chi_{4}\vdd{N'}
+\partial_{+}\chi_{1}\vdd{A_{+}}+\nonumber\\  &   &
-\partial_{+}A_{z}\vdd{\lambda_{1}'}
-\partial_{+}A_{x}\vdd{\lambda_{2}'}
-\partial_{+}A_{-}\vdd{\lambda_{3}'}
-\partial_{+}\lambda_{4}\vdd{D'}\, .
\label{reali}
\end{eqnarray}
A different choice of the preferred $Q_{\alpha}$ would have lead us
to different redefinitions of the fields and different division of
the fields into coordinates of the loop-space and their
differentials.  The relation
\[
Q^{2}_{+}=\dd i_{X_{+}} + i_{X_{+}}
\dd = {\cal L}_{X_{+}}=i\partial_{+}
\]
is the geometric form to express the superalgebra (\ref{salg1}).

Special cases are obtained easily by imposing additional
constraints.  For example, to pick up supersymmetric Maxwell theory
one is to impose reality and a gauge condition:
\[
(C;\chi; M, N; A_{+})=0\, ,
\]
that leaves us with an irreducible multiplet $(A_{-}, A_{x}, A_{z};
\lambda; D)$. Thus we obtain the following relations:
\begin{equation}
{\rm d} = -\lambda_{3}'\vdd{A_{-}}-
\lambda_{2}'\vdd{A_{x}}-\lambda_{1}'\vdd{A_{z}}-
iD'\vdd{\lambda_{4}}\, ,
\end{equation}
\begin{equation}
i_{X} = -\partial_{+}A_{z}\vdd{\lambda_{1}'}
-\partial_{+}A_{x}\vdd{\lambda_{2}'}
-\partial_{+}A_{-}\vdd{\lambda_{3}'}
-\partial_{+}\lambda_{4}\vdd{D'}\, .
\end{equation}
All the statements on the geometric structure of the action can be
now verified.


\vspace*{10mm}
{\bf 4. Nonlinear $\sigma$-model.}  As an example, we shall now
proceed to discuss the two dimensional supersymmetric nonlinear
sigma model. We shall find, that in the quartic fermion term half of
the fermionic degrees of freedom should be interpreted as
coordinates in a superloop space, while the remaining half of the
fermionic degrees of freedom are differentials in this superloop
space. This then identifies the quarctic term as a symplectic two
form.

The action of the $\sigma$-model is
following:
\begin{eqnarray}
{\cal S} & = & {1 \over 2}\int\!{\rm d}^{2}x\{g_{ij}
(\varphi)(\partial_{+}
\varphi^{i}\partial_{-}\varphi^{j}
+\bar{\psi}^{i}i\s{D}\psi^{j}
+\tilde{F}^{i}\tilde{F}^{j})+\label{sigma}\\
 &   & +{1 \over 6}R_{ijkl}(\varphi)\bar{\psi}^{i}\psi^{k}
\bar{\psi}^{j}\psi^{l}\}\, ,\nonumber
\end{eqnarray}
where
\[
\tilde{F}^{i}=F^{i}-{1 \over 2}
\Gamma^{i}_{\, jk}\bar{\psi}^{j}\psi^{k}\, .
\]
It contains a set of real scalar fields $\varphi^{i}$, auxiliary
scalars $F^{i}$ and Majorana spinors $\psi^{i}=(\psi^{i}_{1},
\psi^{i}_{2})$.

As previously, the detailed structure of supersymmetry
transformations suggest a suitable choice for the exterior
derivative and inner multiplication:
\begin{equation}
{\rm d}=\psi_{1}^{i}\vdd{\varphi^{i}}-iF^{i}\vdd{\psi_{2}^{i}}\, ,
\label{dsigma}
\end{equation}
\begin{equation}
i_{X}=i\partial_{-}\varphi^{i}\vdd{\psi_{1}^{i}}
-\partial_{-}\psi_{2}^{i}\vdd{F^{i}}\, ,
\label{isigma}
\end{equation}
with the peculiarity of two dimensions, that one has to make {\sl no
redefinitions} of the fields. Another choice permitted by the
superalgebra would be:
\begin{equation}
{\rm d}'=\psi_{2}^{i}\vdd{\varphi^{i}}+iF^{i}\vdd{\psi_{1}^{i}}\, ,
\nonumber
\end{equation}
\begin{equation}
i_{X'}=i\partial_{+}\varphi^{i}\vdd{\psi_{2}^{i}}
-\partial_{+}\psi_{1}^{i}\vdd{F^{i}}\, .  \nonumber
\end{equation}
The operators $Q={\rm d}+i_{X}$ and $Q'={\rm d}'+i_{X'}$ reproduce
the standard supersymmetry transformations and so the
anticommutation relations of the operators $Q$ and $Q'$ obey the
relations of the supersymmetry algebra (see (\ref{Lie})):
\begin{eqnarray}
QQ'+Q'Q  & = & 0\nonumber\, ,\\
QQ+QQ  & = &  2{\cal L}_{X} =2i\partial_{-}\,,
\label{Q-alg}\\
Q'Q'+Q'Q'  & = &  2{\cal L}_{X'}=2i\partial_{+}\nonumber\, .
\end{eqnarray}
Following the general scheme of section 2 we can find the symplectic
potential corresponding to the action (\ref{sigma}) (if we make the
choice of (\ref{dsigma}) and (\ref{isigma})):
\begin{eqnarray}
\vartheta & = & -\frac{i}{2}g_{ij}\psi_{1}^{i}\partial_{-}\phi^{j}
+\frac{i}{2}g_{ij}\psi_{2}^{i}\partial_{-}F^{j}+\nonumber\\
& &+\frac{1}{2}\Gamma_{i.jk}\psi_{1}^{j}\psi_{2}^{i}\psi_{2}^{k}\, ,
\end{eqnarray}
The action is related to it by
\[
{\cal S}=({\rm d}+i_{X})\vartheta={\cal H}+\omega\, .
\]
One might think that the quartic terms lead to differential forms of
higher degrees than two. In fact, if we write out the relevant terms
in components we obtain the following expression
\be
-{1 \over 2} g_{ik,jl}
\psi_{1}^{i}\psi_{2}^{k}\psi_{1}^{j}\psi_{2}^{l}\label{two-form}\,
,
\ee
that clearly contains the fermionic degrees of freedom
$\psi^{i}_{1}$, that we have identified with differentials,
bilinearly. Hence (\ref{two-form}) is a closed two form in the
superloop space, as expected.


\vspace*{10mm}

{\bf 5.  Conclusions.} We have shown, that the symplectic
interpretation of supersymmetric theories can be based on the
properties of Poincar\'{e} superalgebra and supersymmetry
transformation laws of a general $N=1$ supermultiplet.  Our approach
generalizes the results of \cite{mnp1, mnp2}, where the geometric
structures were discussed for a number concrete models.  In
particular, we have given a model independent definitions of the
exterior derivative in superloop space (\ref{reald}), a contraction
operator $i_X$ with a preferred vector field $X$ (\ref{reali}), and
we have represented a superrotation as a sum of these (\ref{Q}).
The action of supersymmetric models was observed to split into the
sum of a scalar functional (the Hamiltonian) and a two-form (the
symplectic structure), and due to supersymmetry (\ref{hameq}) the
vector field $X$ appears to be Hamiltonian.  In Section 4 we
discussed how the super-loop space formalism applies to the
nonlinear $sigma$-model. In this case, due to the four-fermion term,
in order to have a geometric interpretation it is necessary to use a
superloop space.

\vspace*{7mm}

{\bf Acknowledgements.} K.P. thanks CERN for hospitality and the
Institute of Particle Particle Physics at Helsinki University of
Technology for support. Special thanks are addressed to prof. Antti
Niemi for initiating the study and his keen help at all stages of
this work. The author is also grateful to J. Gracey, P. Howe and V.
Ogievetsky for helpful discussions and advice.



\end{document}